\documentclass[a4paper,11pt]{article}
\pdfoutput=1
\usepackage{jheppub}
\usepackage{multirow}
\usepackage{graphicx}
\usepackage{amssymb, amsmath}
\usepackage{amsfonts}
\usepackage[utf8]{inputenc}
\usepackage{bbold}
\usepackage{bm}
\usepackage{xcolor}
\usepackage{soul}
\usepackage{float}
\usepackage{slashed}
\usepackage{braket}
\usepackage{subcaption}
\newcommand{\ii}{\mathrm{i}}
\newcommand{\rme}{\mathrm{e}}

\newcommand{\parenth}[1]{\left( #1 \right)}
\DeclareMathOperator{\Tr}{Tr}

\makeatletter
\newcommand*{\letterdef@}{}
\newcommand*{\letterdef}[3]{%
	\def\letterdef@##1{\expandafter\newcommand\csname #1\endcsname{#2{##1}}}%
	\@tfor\@tempa :=#3\do{\expandafter\letterdef@\expandafter{\@tempa}}}
\makeatother
\letterdef{c#1} {\mathcal}{ABCDEFGHIJKLMNOPQRSTUVWXYZ} 
\letterdef{rm#1}{\mathrm} {dDmM} 

\newdimen\tableauside\tableauside=1.0ex
\newdimen\tableaurule\tableaurule=0.4pt
\newdimen\tableaustep
\def\phantomhrule#1{\hbox{\vbox to0pt{\hrule height\tableaurule
			width#1\vss}}}
\def\phantomvrule#1{\vbox{\hbox to0pt{\vrule width\tableaurule
			height#1\hss}}}
\def\sqr{\vbox{%
		\phantomhrule\tableaustep
		\hbox{\phantomvrule\tableaustep\kern\tableaustep\phantomvrule\tableaustep}%
		\hbox{\vbox{\phantomhrule\tableauside}\kern-\tableaurule}}}
\def\squares#1{\hbox{\count0=#1\noindent\loop\sqr
		\advance\count0 by-1 \ifnum\count0>0\repeat}}
\def\tableau#1{\vcenter{\offinterlineskip
		\tableaustep=\tableauside\advance\tableaustep by-\tableaurule
		\kern\normallineskip\hbox
		{\kern\normallineskip\vbox
			{\gettableau#1 0 }%
			\kern\normallineskip\kern\tableaurule}%
		\kern\normallineskip\kern\tableaurule}}
\def\gettableau#1 {\ifnum#1=0\let\next=\null\else
	\squares{#1}\let\next=\gettableau\fi\next}
\def\be{\begin{equation}}
\def\ee{\end{equation}}

\title{\boldmath Worldsheet dual of free $\cN=2$ quiver gauge theories}

\author[]{Matthias R.\ Gaberdiel, Francesco Galvagno}
\affiliation[]{Institut f\"ur Theoretische Physik, ETH Z\"urich
\\
	Wolfgang-Pauli-Strasse 27, 8093 Z\"urich, Switzerland
\\}
\emailAdd{gaberdiel@itp.phys.ethz.ch}

\emailAdd{fgalvagno@phys.ethz.ch}

\abstract{For a special family of 4d $\mathcal{N}=2$ superconformal quiver theories, the worldsheet dual corresponding to the free theory is identified. This result is obtained from the recently proposed worldsheet dual of free $\mathcal{N}=4$ SYM by applying a $\mathbb{Z}_k$ orbifold.  In support of our proposal we show that the spectrum of both the untwisted as well as the twisted sectors coincides between the two descriptions. } 


\begin{document}
	\maketitle
	\flushbottom

\section{Introduction}

The prime example of the AdS/CFT correspondence \cite{Maldacena:1997re} relates type IIB string theory on AdS$_5\times {\rm S}^5$ to the maximally supersymmetric $\cN=4$ SYM in 4d \cite{Gubser:1998bc,Witten:1998qj}. In the regime where the AdS radius is large in string units ($\alpha'$ small) and hence can be described in terms of type IIB supergravity, the dual $\cN=4$ theory is strongly coupled. This has allowed one to get insights into strongly coupled gauge theories using supergravity methods. However, exploring the correspondence beyond the supergravity limit has encountered a series of obstacles, mainly due to the difficulties of quantising the full AdS$_5\times {\rm S}^5$ sigma model. 

In \cite{Gaberdiel:2021qbb,Gaberdiel:2021jrv} the tensionless limit ($\alpha^\prime\to \infty$) of string theory on AdS$_5\times {\rm S}^5$ was considered, which is expected to correspond to free $\cN=4$ SYM. Using the twistorial description of AdS$_5\times {\rm S}^5$ of \cite{Berkovits:2004hg} and following the lessons of the worldsheet description of the tensionless string on AdS$_3\times {\rm S}^3 \times \mathbb T^4$ \cite{Eberhardt:2018ouy,Eberhardt:2019ywk,Dei:2020zui}, it was argued that the resulting worldsheet theory reproduces indeed the correct planar spectrum of free $\cN=4$ SYM in 4d. As for the case of AdS$_3\times {\rm S}^3 \times \mathbb T^4$, the worldsheet construction is based on a set of free fields (eight symplectic bosons and four complex fermions, associated to the AdS$_5$ and S$^5$ degrees of freedom, respectively), and they realise the $\cN=4$ superalgebra $\mathfrak{psu}(2,2|4)$ as a current algebra on the worldsheet. 
The other crucial aspect of the construction is that, in addition to the highest weight representations, the worldsheet CFT also contains additional representations that are obtained via spectral flow.
\smallskip

The main goal of this paper is to explore whether a similar description can also be found for gauge theories with less supersymmetry. We shall concentrate on a family of $\cN=2$ superconformal theories which arise as $\mathbb Z_k$ orbifolds of $\cN=4$ SYM. In particular, they can be engineered by considering a stack of D3 branes on a $\mathbb C\times\mathbb C^2/\mathbb Z_k$ orbifold singularity \cite{Douglas:1996sw,Kachru:1998ys,Lawrence:1998ja,Oz:1998hr,Gukov:1998kk,Bertolini:2000dk} with near horizon geometry ${\rm AdS}_5\times {\rm S}^5/\mathbb Z_k$. From the four dimensional gauge theory perspective, the orbifold action breaks the superconformal algebra down to $\mathfrak{su}(2,2|2)$. The theory is therefore best described by reorganising the original $\cN=4$ fields in terms of $\cN=2$ multiplets, and the resulting field content of the $\cN=2$ theory can be represented by a circular quiver, see Figure~\ref{Fig:quiverK}.
Quiver theories have been widely studied from an integrability approach \cite{Beisert:2005he,Gadde:2009dj,Gadde:2010zi,Liendo:2011xb,Pomoni:2011jj,Rey:2010ry,Gadde:2012rv,Pomoni:2013poa,Pomoni:2019oib,Baume:2020ure,Heckman:2020otd,Pomoni:2021pbj}. Recently they have also acquired renewed interest thanks to the power of supersymmetric localisation methods \cite{Pestun:2007rz,Baggio:2014sna,Gerchkovitz:2016gxx,Pini:2017ouj,Fiol:2020ojn,Galvagno:2020cgq,Beccaria:2021ksw,Galvagno:2021bbj}, which lead to some exact results that can be directly compared with supergravity findings \cite{Billo:2021rdb}. As they can be considered the next-to-simplest gauge theories in 4d, they represent a natural class of $\cN=2$ theories for which one may probe the holographic correspondence in the tensionless string regime.

More specifically, we start from the $\cN=4$ proposal of \cite{Gaberdiel:2021qbb,Gaberdiel:2021jrv}, and apply a $\mathbb Z_k$ orbifold to the ${\rm S}^5$ degrees of freedom described by the complex fermions. This breaks the original supersymmetry, while leaving the AdS$_5$ factor untouched; our construction is therefore the natural higher dimensional generalisation of what was considered in \cite{Datta:2017ert,Gaberdiel:2019wjw}. Following the usual orbifold construction of \cite{Dixon:1985jw,Dixon:1986jc} this fixes the worldsheet theory, and we can apply the same prescription as in \cite{Gaberdiel:2021qbb,Gaberdiel:2021jrv} to obtain the physical states. In particular, the single trace operators of the gauge theory of length $w$ should come again from the sector of the worldsheet theory with $w$ units of spectral flow. The main result of our paper is to show that this procedure reproduces correctly the spectrum of the $\cN=2$ quiver gauge theory. While this is in some sense automatic for the untwisted sector where we only need to impose the orbifold projection, the matching of the twisted sectors between the worldsheet description and the dual gauge theory appears non-trivial.

We will pay special attention to the $\mathbb{Z}_2$ orbifold case since it possesses an additional $\mathfrak{su}(2)_L$ global symmetry and is known to be closely related to $\cN=2$ superconformal QCD (SCQCD) which has a single SU$(N)$ gauge group and $N_f=2N$ flavours. 
In particular, the free SCQCD spectrum in the Veneziano limit ($N\to\infty$ while keeping $N_f/N$ fixed) can be obtained from the two-node quiver. 
We will further comment on this point in Section~\ref{sec:concl}. 

\medskip

The paper is organised as follows. In Section~\ref{sec:2} we review the construction of the $\cN=2$ orbifold gauge theories, and in particular describe their spectrum. The exceptional features of the case of the $\mathbb{Z}_2$ orbifold and its relation to SCQCD are explained in Section~\ref{sec:2.1}. Section~\ref{sec:ws} reviews the $\cN=4$ worldsheet theory of \cite{Gaberdiel:2021qbb,Gaberdiel:2021jrv}, and then explains the orbifold theory. In Section~\ref{sec:specmatching} the physical spectrum of the worldsheet theory is determined, and it is shown to agree with that of the $\cN=2$ quiver gauge theory. We also comment on how the special features of the $\mathbb{Z}_2$ orbifold case manifest themselves from the worldsheet perspective. Finally, Section~\ref{sec:concl} contains our conclusions, and there are two appendices where some of the conventions of the $\cN=2$ orbifold gauge theory are explained (Appendix~\ref{App:Rsymmetry}), and the low-lying physical states of the worldsheet theory are worked out explicitly (Appendix~\ref{App:B}).  

\section{$\cN=2$ Orbifolds of $\cN=4$ SYM}\label{sec:2}

We start by reviewing some properties of the $\cN=2$ superconformal theories that arise as orbifolds of $\cN=4$ SYM. After describing the field content and the symmetries of these theories, we describe the orbifold spectrum in terms of untwisted and twisted sectors of the $\cN=4$  theory.


\subsection{The orbifold action}

The $\cN=4$ superconformal symmetry in $4$ dimensions is $\mathfrak{psu}(2,2|4)$. Its bosonic subalgebra is $\mathfrak{so}(2,4)\oplus \mathfrak{su}(4)_R$ (4d conformal algebra and R-symmetry algebra), and the fermionic generators are the $16$ Poincar\'e supercharges $\cQ^A_\alpha$, $\dot{\cQ}^{\dot\alpha}_A$ ($A=1,\dots 4$, $\alpha,\dot\alpha=1,2$), as well as the $16$ conformal supercharges $\cS^A_\alpha$, $\dot{\cS}_{\dot\alpha}^A$. The field content consists of a gauge field $\cA_\mu$, four complex Weyl fermions $\Lambda^A_\alpha$ in the ${\bf 4}$ of the $\mathfrak{su}(4)_R$
R-symmetry algebra, and six scalars $\phi^m$ organised in terms of three complex scalars $\cX,\cY,\cZ$ in the ${\bf 6}$ (antisymmetric self-dual) of $\mathfrak{su}(4)_R$. All the $\cN=4$ fields transform in the adjoint representation of the gauge group, which we take to be ${\rm SU}(Nk)$ for later convenience. 

\begin{figure}[!t]
\begin{center}
\includegraphics[scale=0.6]{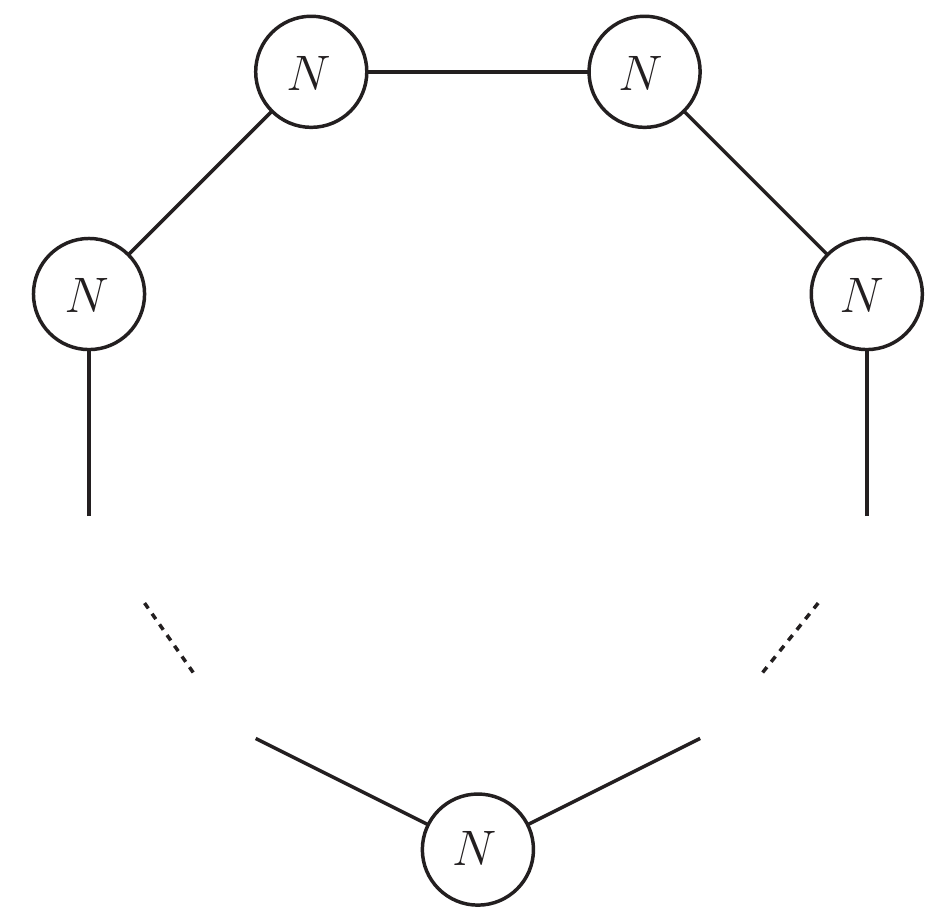}
\end{center}
\caption{The quiver diagram for the $\mathbb Z_k$ orbifold theory. Each node stands for a vector multiplet with $SU(N)$ gauge group, each edge represents a bifundamental hypermultiplet.}
\label{Fig:quiverK}
\end{figure}

The $\mathbb Z_k$ orbifold that we are interested in will reduce the amount of supersymmetry to $\cN=2$, and leads to a superconformal theory with multiple gauge groups, which can be represented as a quiver diagram, see Figure~\ref{Fig:quiverK}. More specifically, we take the fundamental generator of $\mathbb Z_k$ to act as 
\begin{equation}\label{generalMapping}
    \Phi \mapsto \gamma \parenth{\cR\cdot \Phi}\gamma^{-1} \ , 
\end{equation}
where $\cR \in {\rm SU}(4)_R$ is an element of the R-symmetry group, while $\gamma$ is the $kN\times kN$ matrix 
\begin{equation}\label{gammaDef}
    \gamma = \rm{diag} (\mathbb 1_{N\times N}, \, \omega\cdot \mathbb 1_{N\times N} ,\, \cdots, \, \omega^{k-1}\cdot \mathbb 1_{N\times N})\ , \qquad 
    \omega = e^{2\pi \ii / k} \ , 
\end{equation}
acting on the adjoint degrees of freedom of an arbitrary field $\Phi$. 

The choice of $\cR$ determines the supersymmetry breaking, and several R-symmetry breaking patterns are possible, giving rise to different classes of orbifold theories. In order to preserve an $\cN=2$ superconformal symmetry, we consider the subgroup 
\be
{\rm SU}(2)_L \times {\rm SU}(2)_R \times {\rm U}(1)_r \subset {\rm SU}(4)_R \ ,
\ee
and take $\cR$ to be an element of order $k$ in $ {\rm SU}(2)_L$. Since the supercharges transform in the ${\bf 4}$ of the ${\rm SU}(4)_R$, they decompose as ${\bf 2}_L \oplus {\bf 2}_R$, and the orbifold will project out the supercharges associated to ${\bf 2}_L$, and retain those associated to ${\bf 2}_R$. By construction the resulting theory will thus have the R-symmetry group ${\rm SU}(2)_R \times {\rm U}(1)_r$.\footnote{In the special case $k=2$ an additional global symmetry ${\rm SU}(2)_L$ is restored, see subsection \ref{sec:2.1}.}

For this R-symmetry breaking choice, each $\cN=4$ field picks up a phase 
\be\label{Raction}
\cR\cdot \Phi = \omega^{h_\Phi}\, \Phi \ , 
\ee
where $\omega$ is the $k$'th root of unity defined in eq.~(\ref{gammaDef}), and $h_\Phi$ is an integer depending on the ${\rm SU}(4)_R$ charges associated to the field $\Phi$, see appendix \ref{App:Rsymmetry} for further details. In particular the phases are $\Lambda^A\mapsto\parenth{\Lambda^1,\Lambda^2, \omega\Lambda^3,\omega^{-1}\Lambda^4}$ for the Weyl fermions, and $(\cZ,\cX,\cY)\mapsto\parenth{\cZ, \omega \cX,\omega^{-1}\cY}$ for the complex scalars,\footnote{We shall use the calligraphic font $\cX,\cY,\cZ$ to identify the original $\cN=4$ fields, and common uppercase letters for the projections under \eqref{FieldProjection}.} while for the gauge fields (that are ${\rm SU}(4)_R$ singlets) the phases are all trivial, $\cA \mapsto\cA$. This transformation is then combined with the action on the ${\rm SU}(Nk)$ indices, and the invariant fields are those that satisfy 
\begin{equation}\label{constraint}
    \Phi = \omega^{h_\Phi} \, \gamma  \Phi\gamma^{-1} \ , 
\end{equation}
where $\gamma$ is the $Nk\times Nk$ matrix of eq.~(\ref{gammaDef}). Operatively, we may thus implement the orbifold projection by defining
\begin{equation}\label{FieldProjection}
    \Phi =\frac{1}{k} \sum_{\ell=0}^{k-1} \omega^{\ell\,h_\Phi}~\gamma^\ell  \Phi_{(\cN=4)} \gamma^{-\ell} \ ,
\end{equation}
where $\Phi_{(\cN=4)} $ is any ${\cal N}=4$ field. For example, for the gauge fields (for which $h_{\cal A}=0$), the orbifold projection retains the diagonal components only, and hence reduces the gauge group to 
\be\label{quivergauge}
\mathrm{SU}(Nk) \rightarrow \underbrace{\mathrm{SU}(N)^{(1)} \otimes \cdots \otimes \mathrm{SU}(N)^{(k)}}_{\hbox{$k$ copies}} \ , 
\ee
while for the $6$ scalar fields the orbifold invariant combinations are
\begin{align}\label{eq2.8}
    Z = \left( \begin{array}{cccc}
 \varphi^{(1)} &  & &  \\
 &  \varphi^{(2)} & & \\ & &\ddots & \\ & & & ~\varphi^{(k)} \end{array} \right)\ , \quad
     X = \left( \begin{array}{cccc}
 0 & X_{12} & &  \\
 & & X_{23} &  \\ & &\ddots & \ddots \\ X_{k1} & & & 0 \end{array} \right)\ , \quad 
    Y= \left( \begin{array}{cccc}
 0 &  & & Y_{1k} \\
 Y_{21}& 0 & &  \\  & Y_{32} &\ddots & \\  & & \ddots & ~0 \end{array} \right)\ .
\end{align}
Here each $\varphi^{(\ell)}$ is part of an $\cN=2$ vector multiplet transforming in the adjoint representation of $\mathrm{SU}(N)^{(\ell)}$, while the off-diagonal blocks transform in bifundamental representations of the neighbouring gauge groups. Indeed, $X_{\ell\,\ell+1}$ transforms as $\square \times \bar \square$ of $\mathrm{SU}(N)^{(\ell)} \otimes \mathrm{SU}(N)^{(\ell+1)}$, while $Y_{\ell+1\,\ell}$ transforms as $\bar \square \times \square$ of $\mathrm{SU}(N)^{(\ell)} \otimes \mathrm{SU}(N)^{(\ell+1)}$. The analysis for the fermions is similar. 

The overall field content can be represented by a necklace quiver diagram with $k$ nodes, where each node represents an $\mathrm{SU}(N)$ gauge group, see Figure \ref{Fig:quiverK}. The diagonal components then describe $k$ vector multiplets, while the off-diagonal fields give rise to the bifundamental hypermultiplets which represent the conformal matter content of the theory.

\subsubsection{Special case: $\mathbb Z_2$ orbifold}\label{sec:2.1}

\begin{figure}[!t]
\begin{center}
\includegraphics[scale=0.8]{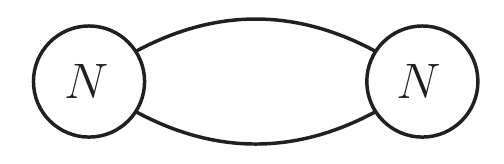}
\end{center}
\caption{The quiver diagram for the $\mathbb Z_2$ orbifold theory. }
\label{Fig:quiver2nodes}
\end{figure}

As we already mentioned before, the case $k=2$ is a bit special since then the surviving group after R-symmetry breaking is actually ${\rm SU}(2)_L \times {\rm SU}(2)_R \times {\rm U}(1)_r$, and hence larger than for $k>2$. The SU$(2)_L$ is an additional global symmetry which plays a crucial role in the determination of the spectrum. To see the enhancement of this symmetry we note that, for $k=2$, the matrix $\gamma$ takes the form 
\be
\gamma =  \left( \begin{array}{cc} \mathbb{1}_{N\times N} & 0 \\ 0 & - \mathbb{1}_{N\times N}  \end{array} \right) \ , 
\ee
and the orbifold invariant fields are 
\begin{equation}
\begin{aligned}  \label{leftcomponents}
Z  &= \left( \begin{array}{cc}
 \varphi^{(1)} & 0 \\
0 &  \varphi^{(2)} \end{array} \right)\ , \quad 
 X  =   \left( \begin{array}{cc}
0 & X_{12} \\
 X_{21} & 0 \end{array} \right)\ , \quad 
 Y  =   \left( \begin{array}{cc}
0 & Y_{12} \\
 Y_{21} & 0 \end{array} \right)\ , \\
\Lambda_I  &=   \left( \begin{array}{cc}
 \lambda_I^{(1)} & 0 \\
0 &  \lambda_I^{(2)} \end{array} \right)\ , \quad \;\;\;
\Lambda_i  =   \left( \begin{array}{cc}
0 & \psi_i^{12} \\\tilde\psi_i^{21} & 0 \end{array} \right)\ , \quad 
A_\mu  =   \left( \begin{array}{cc}
 A_\mu^{(1)} & 0 \\
0 &  A_\mu^{(2)} \end{array} \right)\ , 
 \end{aligned}
\end{equation}
together with their complex conjugates. The gauge group is now broken to the product ${\rm SU}(N)^{(1)}\times {\rm SU}(N)^{(2)}$, and the orbifold theory can be represented by a two-node quiver, see Figure~\ref{Fig:quiver2nodes}.

The different components correspond to the fields of the $\cN=2$ quiver theory and are organised in $\cN=2$ supermultiplets. In particular, the components coming from $Z$, $\Lambda_I$ and $A_\mu$ give rise to two vector multiplets, represented by the two nodes of the quiver. The two bi-fundamental hypermultiplets, represented by the edges in Figure \ref{Fig:quiver2nodes}, are generated by the components coming from $\parenth{X, Y^\dagger,\Lambda_3}$ and $\parenth{Y, X^\dagger,\Lambda_4}$, respectively. For $k=2$, 
these two hypermultiplets are on the same footing, and they are rotated into one another under the 
 additional ${\rm SU}(2)_L$ symmetry.
 
The special features of the $\mathbb Z_2$ orbifold are important in order to understand its relation to superconformal QCD (SCQCD) \cite{Gadde:2009dj,Gadde:2010zi}, which is one of the most studied $\cN=2$ models. SCQCD is a super Yang-Mills theory with a single SU$(N)$ gauge group and $N_f$ fundamental hypermultiplets, where $N_f$ is tuned to $N_f=2N$ in order to preserve conformal invariance. The global symmetry group is U$(N_f)\times\rm{SU}(2)_R\times \rm{U}(1)_r$ (flavour symmetry $\times$ R-symmetry).
In order to understand the relation to SCQCD in more detail, let us start with the two-node quiver gauge theory with gauge group SU$(N)^{(1)}\times$ SU$(N)^{(2)}$ and coupling constants $\lambda_1$ and $\lambda_2$, respectively. If we take the limit $\lambda_2\to 0$  while keeping $\lambda_1=\lambda$ fixed, the second gauge node is decoupled, whereas the rest of the theory reproduces  the SCQCD field content.
At the level of symmetries, the SU$(N)^{(2)}$ group is now a global symmetry and it combines with the SU$(2)_L$ R-symmetry to generate the enhanced U$(N_f=2N)$ flavour symmetry.\footnote{In the quiver gauge theory we only retain states that are singlets with respect to both gauge groups, whereas SCQCD also has states that are not singlet with respect to the flavour symmetry SU$(N)^{(2)}\subset$ U$(N_f=2N)$. The latter states are accounted for by the states in the $\mathbb{Z}_2$ orbifold that are not singlet w.r.t.\ SU$(2)_L$, see \cite{Gadde:2009dj,Gadde:2010zi} for more details.} In our description we are considering the limit in which both $\lambda_1=\lambda_2=\lambda\to 0$ are taken to zero, and hence our analysis makes some contact with the free field limit of SCQCD; we will therefore, throughout the paper, comment on the $\mathbb{Z}_2$ model and its special features, see also Section~\ref{sec:concl}.


\subsection{The orbifold gauge theory spectrum}\label{sec:3} 

Next we want to describe the gauge invariant spectrum of these $\cN=2$ quiver theories. Let us start by reviewing how things work for free $\cN=4$ SYM. The gauge invariant single-trace operators of $\cN=4$ are of the form 
\be\label{letters}
\Tr \parenth{\cS_1\cdots \cS_w}  \ , \qquad \cS_i \in  \left\{ \nabla^s \phi, \nabla^s \Lambda, \nabla^s \bar\Lambda, \nabla^s \cF^+,\nabla^s \cF^- \right\}_{s=0,1,\ldots}\ , 
\ee
where $(\phi^m,\Lambda^A,\bar \Lambda^A, \cF_{\mu\nu})$ describes the field content of the theory, and $\nabla^s= \nabla_{\mu_1} \dots \nabla_{\mu_s}$ denotes (schematically) an $s$-fold covariant derivative. Here the components $\cS_i$ transform in the singleton representation $\mathbb{S}$ of $\mathfrak{psu}(2,2|4)$. The full perturbative spectrum (for the free theory) consists then of the $w$-fold tensor product of the singleton representation $\mathbb{S}$, subject to a cyclicity condition (since we are taking the trace to guarantee the gauge invariance), see \textit{e.g.}\ \cite{Bianchi:2003wx,Beisert:2004di}. 
\smallskip

The gauge-invariant single-trace spectrum of the $\cN=2$ quiver gauge theory can now be described as follows. First of all, instead of $\cS_i\in \mathbb{S}$ running over all states of the singleton representation, we must now restrict $\cS_i$ to the orbifold invariant states $S_i \in \mathbb{S}^{\mathbb{Z}_k}$ that are, for example, obtained from $\cS_i$ upon imposing \eqref{FieldProjection}; this leads to the \textit{untwisted sector} of the orbifold spectrum, that we can schematically describe by 
\begin{equation}\label{untwisted}
    \Tr \parenth{S_1\cdots S_w} \ , \qquad S_i \in  \mathbb{S}^{\mathbb{Z}_k} \ .
\end{equation}
However, the full $\cN=2$ orbifold spectrum is bigger since it also includes \textit{twisted sectors}. Hence it is convenient to introduce a label $\ell$ (with $\ell=0,1,\ldots,k-1$) incorporating both the untwisted sector ($\ell=0$) and the $k-1$ twisted sectors. In general, the states that come from the $\ell$'th twisted sector  are of the form 
\begin{equation}\label{twisted}
    \Tr_{\ell}\parenth{S_1\cdots S_w} \equiv \Tr\parenth{\gamma^\ell \, S_1\cdots S_w} \ , \qquad 
    S_i \in  \mathbb{S}^{\mathbb{Z}_k} \ , 
    \end{equation}
where $\gamma$ is the $kN\times kN$ matrix defined in eq.~\eqref{gammaDef}, which `induces' the twist \cite{Beisert:2005he}. For $\ell=0$, eq.~(\ref{twisted}) obviously reduces to eq.~(\ref{untwisted}).
Note that all of these fields are gauge invariant under the quiver gauge group (\ref{quivergauge}), and that they sit in representations of the $\cN=2$ superconformal algebra $\mathfrak{su}(2,2|2)$. Furthermore since each $S_i \in\mathbb{S}^{\mathbb{Z}_k}$, it follows from eq.~(\ref{constraint}) that 
\begin{equation}\label{gamS}
    \gamma \, S_i = \omega^{-h_i} \, S_i \, \gamma \ ,
\end{equation}
where $h_i$ is the phase associated to $S_i$. Thus if we commute a single $\gamma$ through all the fields in \eqref{twisted}, we conclude that the trace is only non-zero provided that the charges $h_i$ satisfy 
\begin{equation}\label{Zkcharge}
    \frac{1}{k} \, \sum_{i=1}^w h_{i} \in \mathbb Z \ .
\end{equation}
Obviously, this condition also has to hold for the untwisted sector, and in fact, in the untwisted sector the trace of eq.~(\ref{untwisted}) is non-trivial if and only if eq.~(\ref{Zkcharge}) is satisfied. This follows from the structure of the orbifold projection of eq.~(\ref{FieldProjection}), see \textit{e.g.}\ eq.~(\ref{eq2.8}).

The orbifold operators \eqref{twisted} still enjoy an isomorphism with some spin chain states, where the anomalous dimensions can be identified with the eigenvalues of a Hamiltonian $H$, just like in the more familiar $\cN=4$ case. This follows from the fact that the $\cN=4$ $L$-loop Hamiltonian is invariant under the orbifold action provided that we consider operators whose length is longer than $L$ \cite{Beisert:2005he,Solovyov:2007pw}. For the free theory that is of relevance here, we only need to consider the tree-level Hamiltonian, which is therefore automatically orbifold invariant.

\subsubsection{Example: $\Delta=2$ states for $\mathbb Z_{k\geq 3}$ orbifold}\label{sec:exdelta2w3}

To get a sense of what the orbifold spectrum looks like, let us discuss the operators with conformal dimension $\Delta=2$ for the case of the $\mathbb Z_{k\geq 3}$ orbifold. As we mentioned before in Section~\ref{sec:2.1}, the $\mathbb Z_2$ orbifold is somewhat peculiar; we will discuss an example for $k=2$ in Section~\ref{sec:Z2orb} below.

Let us start by describing the $\Delta=2$ operators for $\cN=4$. At $\Delta=2$ all fields are scalars, see eq.~(\ref{Mabmatrix}) in Appendix~\ref{App:Rsymmetry}, and they are of the form 
\begin{equation}\label{N4states}
    \Tr \cZ^2\ , \qquad \Tr \cX^2\ , \qquad \Tr \cZ \cX\ ,\qquad \Tr \cX \cY^\dagger\ , \qquad \cdots \ . 
\end{equation}
Given that there are $6$ scalar fields $(\cX,\cY,\cZ,\cX^\dagger,\cY^\dagger,\cZ^\dagger)$, we get 
20 states from the symmetrised product and 1 from the trace; from the viewpoint of $\cN=4$ they describe the  
$20^\prime$ chiral operators (belonging to the stress tensor multiplet) and the top component of the Konishi multiplet, respectively. In terms of the ${\rm SU}(2)_L \times {\rm SU}(2)_R \times {\rm U}(1)_r$ global symmetry, they decompose as, see  also Appendix~\ref{App:Rsymmetry},
\begin{equation}\label{N4scalarDelta2}
    \begin{split}
        &20^\prime:\quad\quad ~~~\, (\mathbf{1}, \mathbf{1})_{2}~,~~(\mathbf{1}, \mathbf{1})_{-2}~,~~(\mathbf{2}, \mathbf{2})_{1}~,~~(\mathbf{2}, \mathbf{2})_{-1}~,~~ (\mathbf{1}, \mathbf{1})_{0}~,~~(\mathbf{3}, \mathbf{3})_{0}\ ,  \\
        &\text{Konishi}:\quad (\mathbf{1}, \mathbf{1})_{0}\ .
    \end{split}
\end{equation}
In order to obtain the $\cN=2$ orbifold invariant fields we now impose the $\mathbb{Z}_k$ constraint (\ref{Zkcharge}). For $w=2$ and $k\geq 3$ this can only be satisfied by $h_1+h_2=0$, and hence only the combinations 
\begin{equation}
    \label{N4statesResidual}
    \Tr Z^2\ , \qquad \Tr (Z^\dagger)^2\ , \qquad \Tr Z Z^\dagger\ , \qquad \Tr X X^\dagger\ ,\qquad \Tr Y^\dagger Y\ , \qquad \Tr X Y\ ,\qquad \Tr Y^\dagger X^\dagger
\end{equation}
remain. With respect to the residual ${\rm SU}(2)_R \times {\rm U}(1)_r$ R-symmetry, they transform as 
\begin{equation}\label{Z3transformations}
    (\mathbf{1})_{-2}\ , \qquad (\mathbf{1})_{2}\ ,\qquad (\mathbf{1})_{0}\ , \qquad (\mathbf{3})_{0} \ , \qquad (\mathbf{1})_{0} \ . 
\end{equation}
Each of them can now be decomposed into untwisted/twisted combinations by using the definition \eqref{twisted}.  \\ 
For example, for the case of $\Tr X Y$ in a $\mathbb{Z}_3$ orbifold we find 
\begin{align}
&~~~~\Tr_0 X Y=\Tr(\gamma^0 X Y ) = \Tr( X_{12}Y_{21}+X_{23}Y_{32}+X_{31}Y_{13})\ , \notag \\[-.4cm]
 &\nearrow \notag \\[-.2cm]
\Tr X Y& \rightarrow  \Tr_1 X Y=\Tr(\gamma X Y ) =  \Tr( X_{12}Y_{21}+\mathrm{e}^{2\pi\ii/3} X_{23}Y_{32} + \mathrm{e}^{-2\pi\ii/3} X_{31}Y_{13})\ ,  \label{2.20} \\[-.2cm]  &\searrow \notag \\[-.4cm]
&~~~~\Tr_2 X Y=\Tr(\gamma^2 X Y ) = \Tr( X_{12}Y_{21}+\mathrm{e}^{-2\pi\ii/3} X_{23}Y_{32} + \mathrm{e}^{2\pi\ii/3} X_{31}Y_{13})\ . \notag
\end{align}
These expressions are written in terms of $\cN=2$ fields, and they are manifestly $\mathbb Z_3$ eigenstates.
The generalisation to $\mathbb{Z}_k$ works similarly, and we get $k$ (un)twisted sectors, each of them containing states with quantum numbers given in \eqref{Z3transformations}.

Notice that for $\Delta =2$ the number of states in each sector is the same, independently of $\ell$, see eq.~(\ref{2.20}); this is a special feature of this specific case, but not true in general, see \textit{e.g.}\ the following Section.

\subsubsection{States in the $\mathbb Z_2$ orbifold}\label{sec:Z2orb}

As we mentioned before, see Section~\ref{sec:2.1}, the $k=2$ case is somewhat special since the R-symmetry of the quiver gauge theory contains the additional ${\rm SU}(2)_L$ factor. It is therefore instructive to analyse the $\Delta=2$ states also for $k=2$. For $k=2$ in addition to the combinations of (\ref{N4statesResidual}), now also the states with $h_1+h_2=\pm 2$ are possible, \textit{i.e.}\ 
\be
 \Tr X^2 \ , \qquad   \Tr Y^2 \ , \qquad  \Tr (X^\dagger)^2 \ , \qquad   \Tr (Y^\dagger)^2 \ , \qquad \Tr X Y^\dagger \ , \qquad   \Tr X^\dagger Y \ . 
 \ee
Thus there are $13$ potential states, and they all give rise to an untwisted contribution using eq.~\eqref{twisted}. For example for $\Tr X Y^\dagger$ we find 
\begin{align}
&~~~~ \Tr_{0} X Y^\dagger =\Tr( X_{12}Y^\dagger_{12} + Y^\dagger_{21}X_{21})\ , \notag \\[-.4cm]
 &\nearrow \notag \\[-.2cm]
    \Tr X Y^\dagger&  \label{2.22} \\[-.2cm]  &\searrow \notag \\[-.4cm]
&~~~~\Tr_{1} X Y^\dagger =\Tr( X_{12}Y^\dagger_{12} - Y^\dagger_{21}X_{21})\ . \notag
\end{align}
However, for the states associated to $\Tr X^2$, $\Tr (X^\dagger)^2$, $\Tr Y^2$ and $\Tr (Y^\dagger)^2$, the twisted component vanishes (because of the anti-symmetry of the second equation in (\ref{2.22})). Thus altogether we end up with 13 untwisted and 9 twisted states  at level $\Delta =2$. In terms of the ${\rm SU}(2)_L \times {\rm SU}(2)_R \times {\rm U}(1)_r$ global symmetry these states carry the quantum numbers
\begin{equation}\label{UntwistedTwistedDelta2}
    \begin{aligned}
        \rm{untwisted}&:~~~~ (\mathbf{1}, \mathbf{1})_{2}~,~~(\mathbf{1}, \mathbf{1})_{-2}~,~~ 2\times(\mathbf{1}, \mathbf{1})_{0}~,~~(\mathbf{3}, \mathbf{3})_{0}~, \\
        \rm{twisted}&:~~~~(\mathbf{1}, \mathbf{1})_{2}~,~~(\mathbf{1}, \mathbf{1})_{-2}~,~~(\mathbf{1}, \mathbf{1})_{0}~,~~ (\mathbf{3}, \mathbf{1})_{0}~,~~(\mathbf{1}, \mathbf{3})_{0}~.
    \end{aligned}
\end{equation}
The fact that the number of states depends on the twisted sector is not specific to $k=2$; a similar phenomenon arises at higher values of $\Delta$ and $k$.

\section{Worldsheet theory}\label{sec:ws}

In this Section we want to rederive the above results from a worldsheet perspective, generalising the proposal of \cite{Gaberdiel:2021jrv,Gaberdiel:2021qbb} to the orbifold case. We begin by reviewing the salient features of the $\cN=4$ worldsheet theory.  

\subsection{Review of the $\cN=4$ case}

According to the proposal of \cite{Gaberdiel:2021jrv,Gaberdiel:2021qbb} the worldsheet theory dual to free $\cN=4$ SYM in $4$ dimensions consists of eight symplectic boson fields, organised in conjugate pairs as $(\lambda^\alpha,\mu^\dagger_\alpha)$ and $(\lambda^\dagger_{\dot \alpha},\mu^{\dot \alpha})$, where $\alpha,\dot{\alpha}\in\{1,2\}$, together with four pairs of fermions $(\psi^A,\psi^\dagger_A)$, $A=1,\ldots, 4$, 
\begin{equation}\label{freeFieldsComm}
{}[\lambda^\alpha_r, (\mu^\dagger_\beta)_s] = \delta^\alpha_\beta\, \delta_{r,-s} \ , \qquad 
{}[\mu^{\dot{\alpha}}_r, (\lambda^\dagger_{\dot{\beta}})_s] = \delta^{\dot{\alpha}}_{\dot{\beta}}\, \delta_{r,-s} \ , \qquad 
\{ \psi^A_r,(\psi^\dagger_B)_s \} = \delta^A_B\, \delta_{r,-s} \ . 
\end{equation}
These fields give rise to an $\mathfrak{u}(2,2|4)_1$ current algebra. In particular the bosonic subalgebra $\mathfrak{su}(2)\oplus \mathfrak{su}(2)\oplus \mathfrak{su}(4)_R$ is generated by 
\begin{subequations}\label{Bosonalgebra}
\begin{align}
    \cL^\alpha_{\,\beta} &= \mu^\dagger_\beta \lambda^\alpha -\tfrac{1}{2} \delta^\alpha_\beta U~, ~~~~~~~~U=\mu^\dagger_\gamma \lambda^\gamma~,\label{Lab}  \\
    \cL^{\dot\alpha}_{\,\dot\beta} &= \lambda^\dagger_{\dot\beta} \mu^{\dot\alpha} -\tfrac{1}{2} \delta^{\dot\alpha}_{\,\dot\beta} \dot U~, ~~~~~~~~\dot U=\lambda^\dagger_{\dot\gamma} \mu^{\dot\gamma}~,\label{Ldotab} \\
    \cR^A_{\,B} &=\psi^\dagger_B \psi^A - \tfrac{1}{4} \delta^A_B V~, ~~~~~~\, V=\psi^\dagger_C \psi^C~,\label{RAB}
    \end{align}
\end{subequations}
while the off-diagonal generators corresponding to the (conformal) supercharges, translations and special conformal transformations, respectively, are
\begin{subequations}\label{OffDiagalgebra}
\begin{align}
    \cQ^A_{\,\alpha} &= \psi^A\mu^\dagger_\alpha~, ~~~~~~~~ \cS^\alpha_A=\lambda^\alpha\psi^\dagger_A~,\label{Scharges1} \\
    \dot\cQ^{\dot\alpha}_{\,A} &= \mu^{\dot\alpha}\psi^\dagger_A~, ~~~~~~~~ \dot\cS^A_{\dot\alpha}=\psi^A\lambda^\dagger_{\dot\alpha}~, \label{Scharges2} \\
    \cP^{\dot\alpha}_{\,\alpha} &= \mu^{\dot\alpha} \mu^\dagger_{\alpha}~, ~~~~~~~~\cK^\alpha_{\,\dot\alpha} = \lambda^\alpha \lambda^\dagger_{\dot\alpha}~.\label{Translations}
\end{align}
\end{subequations}
It is sometimes convenient to reorganise the generators $U$, $\dot U$, and $V$ as
\begin{equation}\label{Cartans}
    \cB = \tfrac{1}{2}(U+\dot U)~,~~~~~~~~\cC = \tfrac{1}{2}(U+\dot U+V)~,~~~~~~~~ \cD = \tfrac{1}{2}(U-\dot U)~.
\end{equation}
In particular $\cD$ is the dilation operator of the dual $\cN=4$ theory (whose eigenvalue is the conformal dimension $\Delta$), while $\cC$ is central and needs to be removed in order to reduce $\mathfrak{u}(2,2|4)_1$ to $\mathfrak{psu}(2,2|4)_1$; this can be achieved by considering the states that are annihilated by the $\cC_n$ modes with $n\geq 0$. For future reference we note that 
\be\label{Ccharge}
\begin{array}{ll}
{} [ \cC_n, A_r] = \frac{1}{2} A_{n+r} \qquad & \, \hbox{if $A\in \{ \lambda^\dagger_{\dot{\alpha}}, \mu^\dagger_\alpha, \psi^\dagger_A\}$}~, \\[4pt]
{}[ \cC_n, A_r] = - \frac{1}{2} A_{n+r} \quad &\hbox{ if  $A\in \{ \lambda^{\alpha}, \mu^{\dot{\alpha}}, \psi^A\}$\ . }
\end{array}
\ee
Further details can be found in   \cite{Gaberdiel:2021jrv,Gaberdiel:2021qbb}.

\subsubsection{The spectrum in terms of spectrally flowed representations}

The spectrum of the worldsheet theory consists of the NS-sector vacuum representation (in which all of the free fields are half-integer moded), together with the different spectrally flowed images. The ground state of the 
NS sector $\ket{0}$ is characterised by the property that it is annihilated by all the positive modes of the free fields, and as a consequence it is also annihilated by all the $\mathfrak{su}(2)\oplus \mathfrak{su}(2)\oplus \mathfrak{su}(4)_R$ zero modes, as well as by 
\begin{equation}\label{vacuum}
    \cB_0 \ket{0} = \cC_0 \ket{0} = \cD_0 \ket{0} =0~.
\end{equation}
The other representations of the worldsheet theory can be obtained upon spectral flow from this NS-sector vacuum representation. The easiest way to describe these spectrally flowed representations is to consider the NS-sector vacuum representation and to define a modified action on it. More specifically, if we denote by a tilde the modes that act in the usual way on the NS-sector vacuum representation  --- in particular, all the positive tilde modes annihilate $\ket{0}$ --- then the action of the untilde modes that act on this space via 
\begin{equation}\label{spectralFlowmodes}
\begin{aligned}
    (\tilde\lambda^\alpha)_r & = (\lambda^\alpha)_{r-w/2}~,  \qquad  \qquad 
& (\tilde\lambda^\dagger_{\dot\alpha})_r & = (\lambda^\dagger_{\dot\alpha})_{r-w/2}~,   \\ 
(\tilde\mu^{\dot\alpha})_r  & =  (\mu^{\dot\alpha})_{r+w/2} ~, \qquad \qquad 
& (\tilde\mu^\dagger_\alpha)_r  & =  (\mu^\dagger_\alpha)_{r+w/2}  \ ,   \\
(\tilde\psi^I_r) & =  \psi^I_{r-w/2}~, \qquad  \qquad 
& (\tilde\psi^\dagger_{I})_r & =  (\psi^\dagger_I)_{r+w/2}~, \qquad I\in\{1,2\}   \\
(\tilde\psi^i_r) & = \psi^i_{r+w/2}~, \qquad  \qquad
& (\tilde\psi^\dagger_i)_r & = (\psi^\dagger_b)_{r-w/2}~, \qquad i\in\{3,4\}  
\end{aligned}
\end{equation}
define the $w$-fold spectrally flowed representation. In particular, if we denote by $\ket{0}_w$ the ground state with respect to this spectrally flowed action, then the following modes act non-trivially on $\ket{0}_w$
\begin{subequations}\label{VacuumCreat}
\begin{align}
& \mu^{\dot\alpha}_r~, ~~  (\mu^\dagger_\alpha)_r ~, ~~  (\psi^\dagger_I)_r~, ~~   \psi^i_r~,  \qquad\qquad\qquad\qquad\qquad\qquad\qquad\quad~   -\tfrac{w-1}{2} \leq r  \leq \tfrac{w-1}{2} \label{GeneralizedZeroModes} \\
 & 
 \mu^{\dot\alpha}_r~, ~~  (\mu^\dagger_\alpha)_r ~, ~~  (\psi^\dagger_I)_r~, ~~   \psi^i_r~, ~~ \lambda^\alpha_r~,  ~~  (\lambda^\dagger_{\dot\alpha})_r~, ~~ (\psi^I)_r~, ~~  (\psi^\dagger_i)_r ~,  \qquad \qquad\quad  r \leq - \tfrac{w+1}{2}
\end{align}
\end{subequations}
and hence generate the full spectrum from $\ket{0}_w$. For future reference we also note that spectral flow acts as 
\begin{align}
\cD_n & = \tilde{\cD}_n + w\, \delta_{n,0} \ , \qquad \qquad \quad \cR_n = \tilde{\cR}_n + w\, \delta_{n,0} \ , \\
L_n & = \tilde{L}_n + w \, (\cD_n - \cR_n) \ , \qquad  \cC_n = \tilde{\cC}_n \ , 
\end{align}
see \cite{Gaberdiel:2021jrv,Gaberdiel:2021qbb} for more details. In particular, the spectrally flowed ground state $\ket{0}_w$ has the eigenvalues $\cD_0  \, \ket{0}_w = w \, \ket{0}_w$, and $L_0 \, \ket{0}_w = \cC_0 \,\ket{0}_w = 0$.

It was proposed in \cite{Gaberdiel:2021jrv,Gaberdiel:2021qbb} that the physical state condition of the worldsheet string theory removes all of these modes, except for the \textit{wedge modes}, \textit{i.e.}\ the modes in eq.~(\ref{GeneralizedZeroModes}). Furthermore, since they can be thought of as some sort of generalised zero modes, only one copy (from the left- and right-movers) of these wedge modes survives. Finally, on the resulting vector space we need to impose the residual gauge conditions
\be\label{physical}
\cC_n \phi = 0 \ , \ \ n\geq 0 \ , \qquad \qquad (L_0 - pw) \phi = 0 \ , \ \ p\in\mathbb{Z} \ . 
\ee
It was shown in \cite{Gaberdiel:2021jrv,Gaberdiel:2021qbb} that the resulting spectrum reproduces exactly that of the single trace operators of free $\cN=4$ SYM in $4$ dimensions. 

\subsubsection{The Ramond sector and the singleton representation}

The Ramond sector is actually automatically contained in the above description since it can be identified with the image of the NS sector under spectral flow (by one unit). In order to see this we observe from \eqref{GeneralizedZeroModes} that, for $w=1$, the wedge modes are just the zero modes. The mass-shell condition $(L_0 - pw) \phi = 0$ is then trivially satisfied, and the only constraint that needs to be imposed is $\cC_0 \phi=0$; this requires, because of (\ref{Ccharge}), that we apply as many dagger as undagger modes to $\ket{0}_1$. 
The ground state $\ket{0}_1$ (with $\cD_0$ eigenvalue $\cD_0=1$) is the highest weight state of the $\mathfrak{su}(4)_R$ representation $[0,1,0]\cong {\bf 6}$, and the wedge generators of the form $(\psi^\dagger_I)_0 \, \psi^i_0$ are the lowering operators of $\mathfrak{su}(4)_R$ that produce the full $\mathfrak{su}(4)_R$ representation $[0,1,0]$ from the highest weight state. 
The $\cC_0$ neutral combinations of symplectic boson $\mu^{\dot\alpha}_0$ and $(\mu^\dagger_\alpha)_0$ then introduce a non-trivial Lorentz spin, giving rise to the states in the $(\tfrac{s}{2},\tfrac{s}{2};[0,1,0])_{1+s}$ representation of $\mathfrak{su}(2) \oplus \mathfrak{su}(2) \oplus \mathfrak{su}(4)_R$ where $s=0,1,\ldots$, and the subscript denotes the eigenvalue with respect to $\cD_0$, which corresponds to the conformal dimension $\Delta$ in the dual gauge picture. 

On the other hand, applying the fermionic generators $(\psi^\dagger_I)_0\, \mu^{\dot{\alpha}}_0$ and $(\mu^\dagger_\alpha)_0 \,\psi^i_0$ to $\ket{0}_w$ gives rise to states that transform as $(0,\tfrac{1}{2} ; [1,0,0])_{3/2}$ and $(\tfrac{1}{2},0 ; [0,0,1])_{3/2}$, respectively, and, as before, the $\cC_0$ neutral symplectic boson generators then produce an infinite tower of states with increasing $\cD_0$ eigenvalue. Continuing in this manner (and using that because of the Fermi statistics of the fermionic zero modes we can only apply each of the generators $(\psi^\dagger_I)_0\, \mu^{\dot{\alpha}}_0$ and $(\mu^\dagger_\alpha)_0 \,\psi^i_0$ at most once), we find for the full spectrum 
\begin{equation}
    \begin{aligned}\label{2Dsingleton}
\cR_0 =  \bigoplus_{s=0}^\infty  &\bigg[ \parenth{ \tfrac{s}{2},\tfrac{s}{2};[0,1,0]}_{1+s}  \oplus \parenth{ \tfrac{s}{2},\tfrac{s+1}{2} ; [1,0,0]}_{\frac{3}{2}+s}
\oplus \parenth{\tfrac{s+1}{2},\tfrac{s}{2}; [0,0,1] }_{\frac{3}{2}+s}     \\
& \oplus 
\parenth{ \tfrac{s}{2}+1,\tfrac{s}{2} ; [0,0,0]}_{2+s} \oplus \parenth{ \tfrac{s}{2},\tfrac{s}{2}+1 ; [0,0,0]}_{2+s}
 \bigg] \ . 
\end{aligned}
\end{equation}
This is the so-called singleton representation of $\mathfrak{psu}(2,2|4)$, and the different terms in the sum account precisely for the different letters in $\mathbb{S}$, see eq.~(\ref{letters}).

For $w>1$, on the other hand, we have effectively $w$ copies of these zero mode wedge generators, and the $\cC_n\phi=0$ condition for $n\geq 0$ ensures that the resulting states lie in the $w$'th tensor power of the above singleton representation.\footnote{Strictly speaking, unlike what was claimed in \cite{Gaberdiel:2021qbb}, this does not directly follow from $\cC_n\phi=0$ for $n\geq 0$ since the modes that appear in  \cite[eq.~(4.10)]{Gaberdiel:2021qbb} should also include those combination of wedge modes that sum up to $n$ mod $w$. However, we have checked experimentally that this distinction seems to be immaterial.} The mass-shell condition $(L_0 - pw) \phi = 0$ then implies that we only retain the cyclically invariant states in this $w$-fold tensor product, and this precisely accounts for the single-trace states of $\cN=4$ SYM, see eq.~(\ref{letters}).

\subsection{The Orbifold action on the worldsheet}\label{sec:3.2}

We now want to identify the orbifold action on the dual AdS background in the worldsheet description. Since $\mathbb{Z}_k$ only acts on the  ${\rm S}^5$ factor which is described by the worldsheet fermions, we should expect that we only need to orbifold the fermions, but leave the symplectic bosons invariant. We propose that the primitive generator  $g\in\mathbb Z_k$ acts as
\begin{equation}\label{worldsheetOrbaction}
    g(\psi^I) = \psi^I~, \qquad g(\psi^3)=\omega\, \psi^3~,\qquad g(\psi^4)=\omega^{-1} \psi^4~, \qquad\qquad \omega=\text{e}^{2\pi\ii/k}~,
\end{equation}
and correspondingly $g(\psi^\dagger_I) = \psi^\dagger_I$,  $g(\psi^\dagger_3) = \omega^{-1} \psi^\dagger_3$, and $g(\psi^\dagger_4) =  \omega\, \psi^\dagger_4$.

These phase factors only leave the super(conformal) generators of \eqref{Scharges1} and \eqref{Scharges2} invariant provided that the index takes the values $I=1,2$, while the R-symmetry generators \eqref{RAB} are reduced to $\cR^I_J = \psi^\dagger_J \psi^I$, parametrising the residual $\mathfrak{su}(2)_R \oplus \mathfrak{u}(1)_r$ R-symmetry. In particular, the $\cN=2$ superconformal algebra $\mathfrak{su}(2,2|2)$ is preserved.\footnote{A similar realisation of the $\cN=2$ superconformal algebra was used in \cite{Liendo:2011xb} to compute the one-loop Hamiltonian for SCQCD. We thank Elli Pomoni for a discussion about this.} 
We also note that the orbifold action \eqref{worldsheetOrbaction} leaves the $V$ generator of eq.~(\ref{RAB}) invariant, and hence also the `central' generator $\cC$ is untouched (see eq.~\eqref{Cartans}). As a consequence, the coset condition $\cC_n \phi=0$ is unaffected by the orbifold. Obviously, the same is also true for the mass-shell condition. 

\subsubsection{Untwisted sector}

The untwisted sector of the worldsheet theory now consists of all states of the above (unorbifolded) worldsheet theory, subject to the condition that they are orbifold invariant. As we explained before, the physical states in the $w$'th flowed sector consist of the cyclically invariant states in the $w$-fold tensor product of the singleton representation. At the level of the singleton representation, it is easy to see how the orbifold group acts: decomposing the $\mathfrak{su}(4)_R$ representations in terms of  $\mathfrak{su}(2)_R \oplus \mathfrak{u}(1)_r$ representations as was done in Section~\ref{sec:2} (see in particular the discussion after eq.~\eqref{Raction}) we find 
\begin{align}\label{RsymmDecomposition}
    [0,1,0] &\mapsto [0]_{1} + [0]_{-1}+ \omega \left[\tfrac{1}{2} \right]_0+ \omega^{-1} \left[\tfrac{1}{2} \right]_0~, \notag \\[4pt]
    [1,0,0] &\mapsto \left[\tfrac{1}{2} \right]_{-1/2}+ \omega \left[0 \right]_{1/2}+ \omega^{-1} \left[0\right]_{-1/2}~,  \\
    [0,0,1] &\mapsto \left[\tfrac{1}{2} \right]_{1/2}+ \omega^{-1} \left[0 \right]_{1/2}+ \omega \left[0\right]_{-1/2}~\ , \notag 
\end{align}
where the subscript denotes the $\mathfrak{u}(1)_r$ charge. This orbifold action therefore reproduces precisely the $\mathbb{Z}_k$ charges from the action in eq.~(\ref{Raction}), see Appendix~\ref{App:Rsymmetry}. 

Thus the untwisted sector states on the worldsheet consists of those cyclically invariant combinations of singleton fields that satisfy (\ref{Zkcharge}). As is explained in the paragraph below that equation, this then reproduces precisely the untwisted sector of the field theory.

\subsubsection{Twisted sectors}

In addition to the untwisted sector the worldsheet theory also has $k-1$ twisted sectors in which the fermions (that are not invariant under the orbifold action) will have fractional modes. Mirroring what we did for the gauge theory, we label the different sectors of the worldsheet theory by $\ell$, where in the $\ell$'th twisted sector the 
twisted modes (before spectral flow) have the mode numbers
\begin{equation}\label{twistedFractional}
    \psi^3_{r+\frac{\ell}{k}}~,~~~\psi^4_{r-\frac{\ell}{k}}~,~~~ (\psi^\dagger_3)_{r-\frac{\ell}{k}}~,~~~ (\psi^\dagger_4)_{r+\frac{\ell}{k}}~,~~~~~~~~r\in \mathbb Z+\tfrac{1}{2} \ . 
\end{equation}
(Thus $\ell=0$ describes the untwisted sector.) In the following it will be convenient to take $\ell$ to lie in the range $-\lfloor\tfrac{k}{2}\rfloor\leq \ell <\lfloor\tfrac{k}{2}\rfloor$, so that $|\tfrac{\ell}{k}|\leq \tfrac{1}{2}$. Spectral flow also acts on these `twisted' modes as before, and it is again natural to postulate that, after imposing the physical state condition, the surviving wedge modes will be 
\begin{equation}\label{(un)twistedwedgemodes}
    \begin{aligned}
    &\mathrm{untwisted~modes:} \qquad  \mu^{\dot\alpha}_r~, \quad  (\mu^\dagger_\alpha)_r ~, \quad  (\psi^\dagger_I)_r~,  \quad \qquad  &-\tfrac{w-1}{2}\leq r \leq \tfrac{w-1}{2} \ \phantom{.}\\
    &\mathrm{twisted~modes:}\quad \qquad~\psi^3_{r+\frac{\ell}{k}}~, \quad \psi^4_{r-\frac{\ell}{k}}  \quad \qquad & -\tfrac{w-1}{2}\leq r \leq \tfrac{w-1}{2} \ .
\end{aligned}
\end{equation}
All of them act non-trivially on the twisted sector ground state $\ket{0}_w$, and on the resulting Fock space we then need to impose the residual gauge conditions of eq.~(\ref{physical}). 

We should mention that the shift of the (fractional) mode numbers in eq.~\eqref{(un)twistedwedgemodes} is only well motivated for $k>2$ since for $k=2$ there is no difference between the shift for $\psi^3$ and $\psi^4$ in the $\ell=1$ twisted sector. Thus the assignement of the wedge modes for $k=2$ is ambiguous, and in fact the correct prescription will differ from eq.~\eqref{(un)twistedwedgemodes} for $k=2$;\footnote{As we shall see, this subtlety is also at the root of the symmetry enhancement of the R-symmetry group, see Section \ref{sec:2.1}.}  we will come back to case $k=2$ in more detail in Section~\ref{sec:4.2}.

\section{Physical spectrum and spin chain picture}\label{sec:specmatching}

Next we want to determine the physical spectrum for these different twisted sectors, \textit{i.e.}\ find the states in the Fock space generated by the wedge modes in eq.~(\ref{(un)twistedwedgemodes}) that satisfy the residual physical gauge conditions of eq.~(\ref{physical}). We have worked this out explicitly for a few low-lying cases, see below, but one can actually also give a more abstract argument. To do so we go to a position basis for the wedge modes following \cite{Gaberdiel:2021qbb}: for the untwisted wedge modes we define as before 
\be\label{disF}
\hat{\Phi}_j = \frac{1}{\sqrt{w}}\, \sum_{r=- (w-1)/2}^{(w-1)/2} \Phi_r \, e^{-2\pi \ii\frac{rj}{w}}  \ , 
\ee
while for the twisted wedge modes in the $\ell$'th twisted sector we define instead 
\be\label{distF}
\hat{\Phi}^{({\rm twis})}_j = \frac{1}{\sqrt{w}}\, \sum_{r=- (w-1)/2}^{(w-1)/2} \Phi_{r\pm \frac{\ell}{k}} \, e^{-2\pi \ii\frac{rj}{w}}  \ , 
\ee
where the sign depends on whether we consider $\Phi=\psi^3$ ($+$) or $\Phi=\psi^4$ ($-$).
For the analysis of the $\cC_n=0$ condition the shift in the mode numbers is immaterial, and thus we end up  (after imposing the $\cC_n=0$ condition with $n\geq 0$) with the $w$'th tensor product of the singleton representation. However, the periodicity condition is now modified since for a twisted mode, 
\begin{align}
[e^{\frac{2\pi \ii}{w}\, L_0}, \hat{\Phi}^{({\rm twis})}_j \, ]  & =   \frac{1}{\sqrt{w}}\, \sum_{r=- (w-1)/2}^{(w-1)/2} 
[e^{\frac{2\pi \ii}{w}\, L_0}, \Phi_{r\pm \frac{\ell}{k}}] \, e^{-2\pi \ii\frac{rj}{w}} \notag  \\
& =   \frac{1}{\sqrt{w}}\, \sum_{r=- (w-1)/2}^{(w-1)/2} 
\Phi_{r \pm \frac{\ell}{k}} \, e^{- \frac{2\pi \ii}{w}\,(r\pm \frac{\ell}{k})} \, e^{-2\pi \ii\frac{rj}{w}} \\ 
& =  e^{ \mp \frac{2\pi \ii \ell}{k w}} \,  \frac{1}{\sqrt{w}}\, \sum_{r=- (w-1)/2}^{(w-1)/2} 
\Phi_{r\pm \frac{\ell}{k}} \, e^{-2\pi \ii\frac{r(j+1)}{w}} = e^{ \mp \frac{2\pi \ii \ell}{k w}} \, \hat{\Phi}^{({\rm twis})}_{j+1}  \notag ~.
\label{phase}
\end{align}
This is to be contrasted with the corresponding equation for an untwisted mode, see \cite[eq.~(4.6)]{Gaberdiel:2021qbb}
\be
[e^{\frac{2\pi \ii}{w}\, L_0}, \hat{\Phi}^{({\rm untwis})}_j \, ]   = \hat{\Phi}^{({\rm untwis})}_{j+1} \ ,
\ee 
and thus we can write the general case as 
\be\label{onestep}
[e^{\frac{2\pi \ii}{w}\, L_0}, \hat{\Phi}_j \, ]   = e^{ - \frac{2\pi \ii \ell}{k w} h_\Phi } \, \hat{\Phi}_{j+1} \ , \qquad 
g(\Phi) = \omega^{\, h_\Phi} \, \Phi \ ,
\ee
where $h_\Phi$ is the orbifold charge, see eq.~(\ref{worldsheetOrbaction}), and $\ell$ denotes the twisted sector. We also need to impose the orbifold invariance condition in the twisted sector, and this amounts to demanding that the total orbifold charge vanishes mod $kw$, \textit{i.e.}\ that the number of $\psi^3$ modes equals the number of $\psi^4$ modes mod $kw$. Then the product of phases on the right-hand-side of (\ref{onestep}) is trivial, and the mass-shell condition, \textit{i.e.}\ the requirement that the $L_0$ charge is a multiple of $w$, implies again that we only retain the cyclically invariant combinations of the $w$'th fold tensor product of the singleton. However, now the individual fields pick up a phase upon translation, and if we move a field once around the entire spin chain, \textit{i.e.}\ if we consider 
\be\label{loop}
[e^{2\pi \ii\, L_0}, \hat{\Phi}_j \, ] = e^{-  \frac{2\pi \ii \ell}{k } h_\Phi} \, \hat{\Phi}_{j}~,
\ee 
the phase depends on the orbifold charge. This then reproduces precisely the gauge theory result, 
see eq.~(\ref{twisted}), since the analogue of moving the field around the trace is to move it past the $\gamma^\ell$ term, for which, because of eq.~(\ref{gamS}), we find 
\be
\gamma^{\ell}\, S_j \, \gamma^{-\ell} = e^{- \frac{2\pi \ii  \ell}{k} h_j}  \, S_j \ .
\ee
Since the orbifold charge agrees precisely with $h_j$, this then reproduces eq.~(\ref{loop}). We have also checked this agreement explicitly for the $\mathbb{Z}_3$ case for the states with $\Delta=2, \tfrac{5}{2}, 3$. 


\subsection{DDF-like operators}

As in \cite{Gaberdiel:2021jrv}, one can actually explicitly construct the physical spectrum by introducing a family of DDF-like operators \cite{DELGIUDICE1972378} which create physical states from the spectrally flowed ground state and thereby generate the full spectrum. In the $\cN=4$ case these operators were defined for $0\leq m \leq w-1$ via 
\begin{equation}\label{DDFold}
(\cS_{U}^{~V})_{m} = \!\! \sum_{r =  m - \frac{w-1}{2}}^{\frac{w-1}{2}} \!\! (\cY_U)_{r} \, (\cZ^V)_{m-r} \ ,
\end{equation}
where $\cY_U= (\mu^\dagger_{\alpha},\lambda^\dagger_{\dot{\alpha}},\psi^\dagger_A)$ and $\cZ^U = (\lambda^{\alpha},\mu^{\dot{\alpha}},\psi^A)$ collectively denote all the worldsheet fields.

In the present context we can still define the analogue of (\ref{DDFold}) provided that both operators are unaffected by the orbifold action, 
\be\label{DDFopsU}
({\cal U}_{M}^{~N})_{m} = \!\! \sum_{r =  m - \frac{w-1}{2}}^{\frac{w-1}{2}} \!\! (Y_M)_{r} \, (Z^N)_{m-r} \ , 
\ee
where $Y_M= (\mu^\dagger_{\alpha},\lambda^\dagger_{\dot{\alpha}},\psi^\dagger_I)$ and $Z^N = (\lambda^{\alpha},\mu^{\dot{\alpha}},\psi^I)$ only involve the fields that are invariant under $g$. In addition, we now have `twisted' DDF operators of the form 
\begin{align}\label{DDFopsT}
({\cal T}_{M}^{~v})_{m\pm\frac{\ell}{k}} & = \!\!\sum_{r =  m - \frac{w-1}{2}}^{\frac{w-1}{2}}\!\! (Y_M)_{r} \, (X^v)_{m-r\pm\frac{\ell}{k}}   \ , 
\end{align}
where $X^v=(\psi^i)$, and the shift in the mode number of eq. \eqref{DDFopsT} depends on the choice of the twisted oscillator: the $\psi^3$ generated is shifted as $+\tfrac{\ell}{k}$, while $\psi^4$ picks up $-\tfrac{\ell}{k}$, see eq.~\eqref{(un)twistedwedgemodes}.\footnote{There are obviously also similar twisted operators where we consider the $\psi^\dagger_{3,4}$ modes, but since the latter are annihilation operators, we do not need these generators for the creation of the physical states.} Provided that $m\geq 0$ these generators commute with ${\cal C}_n$ with $n\geq 0$ on the wedge Fock space, and thus map physical states to physical states, except that we still need to impose the orbifold and mass-shell condition. We have checked that these DDF operators are sufficient to generate the full physical spectrum; some low-lying examples are spelled out in Appendix~\ref{App:B}. 

We should also note that the zero modes of (\ref{DDFopsU}) form the Lie algebra of $\mathfrak{su}(2,2|2)$, which is the appropriate symmetry of an ${\cal N}=2$ theory. Thus the resulting physical spectrum manifestly has this symmetry. 


\subsection{$\mathbb{Z}_2$ orbifold and symmetry enhancement}\label{sec:4.2}
As in the gauge theory set up, the $\mathbb{Z}_2$ case is special, due to  the additional $\mathfrak{su}(2)_L$ global symmetry, see Section \ref{sec:2.1}. Here we want to explain how this manifests itself from the worldsheet perspective. As we mentioned before, see the last paragraph of Section~\ref{sec:ws}, for $k=2$ the twisted modes in the $\ell=1$ sector are of the form 
\begin{equation}\label{twistedZ2}
\psi^i_{r+\frac{1}{2}} \qquad (\psi^\dagger_i)_{r+\frac{1}{2}}\ ,
\end{equation}
and thus there is no difference between how $\psi^3$ and $\psi^4$ are twisted. As a consequence it is a bit unnatural to take the non-trivial wedge modes to be different as in eq.~(\ref{(un)twistedwedgemodes}). For $k=2$ we therefore postulate that the spectrum generating wedge modes in the $\ell=1$ sector take the form 
\begin{equation}\label{wedgemodesZ2}
k=2 \ , \ \ \ell=1: \qquad     \mu^{\dot\alpha}_r~, \quad  (\mu^\dagger_\alpha)_r ~, \quad  (\psi^\dagger_I)_r~, \quad \psi^i_{r+\frac{1}{2}}  \qquad  -\tfrac{w-1}{2}\leq r \leq \tfrac{w-1}{2} \ .
\end{equation}
As a consequence, this allows us to define the additional DDF modes 
\be
(U_{u}^{\prime \, v})_{m} = \!\!\sum_{r =  m - \frac{w-1}{2}}^{\frac{w-1}{2}}\!\! (W_u)_{r-\frac{1}{2}} \, (X^v)_{m+\frac{1}{2}-r} \ , \label{DDFopsUZ2} 
\ee
where $W_u= (\psi^\dagger_i)$ and, as before, $X^v=(\psi^i)$. In particular, the corresponding zero modes define the Lie algebra of $\mathfrak{su}(2)$, and this enhances the $\mathfrak{su}(2,2|2)$ symmetry by $\mathfrak{su}(2)_L$, mirroring exactly what we obtained also from the gauge theory side, see Section~\ref{sec:2.1}. 

We have also checked explicitly for a few low-lying cases ($\Delta=2, \tfrac{5}{2}, 3$) that the above prescription reproduces the correct gauge theory spectrum, see Appendix~\ref{App:B}.



\section{Conclusion and future perspectives}\label{sec:concl}

In this paper we have made a proposal for the worldsheet theory that is dual to a family of 4d $\cN=2$  superconformal quiver theories in the free limit. The basic idea was to consider a $\mathbb Z_k$ orbifold of the proposed duality between $\cN=4$ SYM in 4d and the free field worldsheet theory of \cite{Gaberdiel:2021qbb,Gaberdiel:2021jrv}. More specifically, we considered a $\mathbb Z_k$ orbifold of $\cN=4$ SYM that preserves the $\cN=2$ superconformal symmetry, and then translated this action to the worldsheet variables. This allowed us to identify the worldsheet orbifold --- the orbifold only affects the  complex worldsheet fermions that account for the S$^5$ degrees of freedom --- and thus to specify the worldsheet theory completely. The main piece of evidence for our proposal is that the physical spectrum of this worldsheet theory reproduces the spectrum of the $\cN=2$  superconformal field theory. While this is more or less obvious (by construction) for the untwisted degrees of freedom, the matching of the twisted sectors is non-trivial.


The natural next step would be to try to reproduce the correlation functions of the 4d $\cN=2$  superconformal quiver theory from the worldsheet perspective. For the case of string theory on  AdS$_3\times {\rm S}^3 \times \mathbb T^4$, this was done in \cite{Eberhardt:2018ouy,Dei:2020zui}, see also \cite{Eberhardt:2020akk,Knighton:2020kuh} for the generalisation to higher genus, and it would be very interesting to see how this generalises to the present case (or indeed the $\cN=4$ case of \cite{Gaberdiel:2021qbb,Gaberdiel:2021jrv}). It would also be very interesting to see how (and whether) the simplification that occurs for operators of large length ($w$ large), see \cite{Gaberdiel:2020ycd}, manifests itself for the 4d setup. 

Another natural direction is to generalise our analysis to other kinds of orbifolds \cite{Kachru:1998ys,Oz:1998hr}, preserving different amounts of supersymmetry,\footnote{For example, the $\mathbb Z_3$ orbifold theory described in \cite{Kachru:1998ys} has a R-symmetry breaking pattern such that the orbifold acts as $(\cZ,\cX,\cY)\mapsto (\rme^{2\pi\ii/3}\cZ,\rme^{2\pi\ii/3}\cX,\rme^{2\pi\ii/3}\cY)$, preserving an $\cN=1$ supersymmetry.} or leading even to non supersymmetric theories. It would be very interesting to study how the worldsheet construction would be affected for these cases. 

Furthermore, the connection of the $\cN=2$ $\mathbb{Z}_2$ orbifold to SCQCD, see Section~\ref{sec:2.1}, is worth to be explored in more detail. The holographic properties of SCQCD are still pretty elusive,\footnote{In \cite{Gadde:2009dj} a seven dimensional subcritical string background containing an ${\rm AdS}_5\times {\rm S}^1$ factor was proposed as the dual geometry of SCQCD for $\lambda\gg 1$.} but our worldsheet formulation may provide some insights about the holographic properties of $\cN=2$ SCQCD in the dual tensionless stringy regime. 
In particular, our worldsheet analysis describes the $\mathbb Z_2$ orbifold point ($\lambda_1=\lambda_2=\lambda$) in the free limit $\lambda\to 0$. This is related to the Veneziano limit ($N\to\infty$ with $N_f/N$ fixed) of SCQCD, since the SU$(2)_L$ singlet part of the $\mathbb Z_2$ spectrum (both in the twisted and untwisted sectors) can be directly decomposed in terms of SCQCD plus a decoupled vector multiplet, see \cite{Gadde:2010zi} for more details.\footnote{For example in the list \eqref{UntwistedTwistedDelta2} for the $\Delta=2$ spectrum, three states $(\mathbf{1}, \mathbf{1})_{2}$, $(\mathbf{1}, \mathbf{1})_{-2}$, $(\mathbf{1}, \mathbf{1})_{0}$ are part of the decoupled vector multiplet, while the remaining SU$(2)_L$-singlets states (from both the untwisted and twisted sectors) directly translate into the SCQCD spectrum as U$(N_f)$ singlets. This mapping fails for the SU$(2)_L$ non-singlets $(\mathbf{3}, \mathbf{3})_{0}$ and $(\mathbf{3}, \mathbf{1})_{0}$, which should be combined to form a SU$(2)_R$ triplet with open U$(N_f)$ indices, and hence cannot be immediately recovered from the orbifold construction.} 

It would obviously also be interesting to study the behaviour of our duality upon turning on the marginal Yang-Mills coupling $\lambda$, but before studying this problem for our $\cN=2$ setup, it would be useful to analyse first the simpler case of AdS$_3\times {\rm S}^3 \times \mathbb T^4$ \cite{progress}.

\vskip 1.5cm
\noindent {\large {\bf Acknowledgments}}
\vskip 0.2cm
We thank S.~Benvenuti, L.~Bianchi, M.~Bill\`o, G.~Cuomo, M.~Frau, R.~Gopakumar, S.~Komatsu, B.~Knighton, A.~Lerda, L.~Lin, A.~Pini, E.~Pomoni, J.~Russo and J.~Vosmera
for many useful discussions and suggestions.

\noindent
This work was supported by the Swiss National Science Foundation through a personal grant
and via the NCCR SwissMAP. 
\vskip 1cm

\begin{appendix}

\section{R-symmetry decomposition}\label{App:Rsymmetry}

In this appendix we derive the phases that appear in the $\mathbb{Z}_k$ action defined by (\ref{Raction}). Since $\cR \subset {\rm SU}(4)_R$, these phases can be directly read off from the ${\rm SU}(4)_R$ transformation properties of the fields. 

First of all, the gauge fields $\cA_\mu$ are ${\rm SU}(4)_R$ invariant, and hence their associated phases $h_A=0$ are trivial. The Weyl fermions transform as a vector under ${\rm SU}(4)_R$ --- they transform in the fundamental ${\bf 4}$  --- while the scalars sit in the ${\bf 6}$, and hence can be described by an antisymmetric matrix $M_{AB}$ ($A,B=1,\dots 4$),
 \begin{equation}\label{Mabmatrix}
\Lambda^{A}=\parenth{\begin{array}{c}
\Lambda^{1}\\\Lambda^{2}\\\Lambda^{3}\\\Lambda^{4}\end{array}}~,~~~~~~
M_{AB}=\parenth{\begin{array}{cc|cc}
0 & \cZ & \cX & \cY\\
-\cZ & 0 &  \cY^\dagger & - \cX^\dagger\\
\hline -\cX & - \cY^\dagger & 0 &  \cZ^\dagger\\
-\cY & \cX^\dagger & - \cZ^\dagger & 0\end{array}}\ , 
\end{equation}
where $\cR\in {\rm SU}(4)_R$ acts on $\Lambda^A$ from the right, while $M_{AB}$ transforms as $M \rightarrow \cR^{-1} \, M \, \cR$. 

The $\cR$ generators that are relevant for (\ref{Raction}) all sit in the subgroup ${\rm SU}(2)_L \subset {\rm SU}(4)_R$, which we may choose to be 
\be
\begin{array}{cc}
1 \\
2 \\
3  \\
4 \end{array}
\parenth{\begin{array}{cc|cc}
{\rm SU}(2)_R  \times {\rm U}(1)_r&  & \,\\
 &  & \,\\
\hline  &  & \,\\
 &  & \, & {\rm SU}(2)_{L} \times {\rm U}(1)_r^* 
 \end{array}} \ \subset {\rm SU}(4)_R \ . 
 \ee
Thus if we organise the $A,B$ vector indices into two groups as $I, J = 1,2$ and $i,j = 3,4$, then $I,J$ are vector indices with respect to ${\rm SU}(2)_R$, while $i,j$ are vector indices with respect to ${\rm SU}(2)_L$. It is then immediate to read off how the different fields transform under $\cR$. 

First of all, the scalar $\cZ$ and its conjugate $\cZ^\dagger$ are both ${\rm SU}(2)_L \times {\rm SU}(2)_R$ singlets, but they carry (opposite) charge with respect to ${\rm U}(1)_r$. On the other hand, the scalars 
$\cX$ and $\cY$ (as well as their conjugates $\cX^\dagger$ and $\cY^\dagger$) sit in the off-diagonal part $M_{Ii}$ of the matrix \eqref{Mabmatrix}, and hence transform as $(\bf 2,\bf 2)$ under ${\rm SU}(2)_L \times {\rm SU}(2)_R$ but are neutral under ${\rm U}(1)_r$. Thus they pick up phases $\omega$ and $\omega^{-1}$, respectively. Finally, the Weyl fermions decompose as $\lambda_I$ in $(\bf 1,\bf 2)$ and $\lambda_{i}$ in $(\bf 2,\bf 1)$ with respect to ${\rm SU}(2)_L \times {\rm SU}(2)_R$; thus the $\lambda_I$ transform trivially under $\cR$, while the two components of $\lambda_i$ pick up the phases $\omega$ and $\omega^{-1}$, respectively. 

\section{Explicit worldsheet spectrum}\label{App:B}

In this appendix we give some details about the worldsheet description of some low-lying states of the orbifold theory. 

\subsection{The $w=2$ states at $\Delta=2$}

The gauge theory states at $w=2$ and $\Delta=2$ were enumerated in Section~\ref{sec:exdelta2w3}, and it was analysed which states survive for the $\mathbb{Z}_2$ and $\mathbb{Z}_3$  orbifold, respectively. In this appendix we want to reproduce these results from the worldsheet perspective. 

In the $w=2$ sector, the ground state has $\cD_0 \, \ket{0}_2 = 2 \, \ket{0}_2$, so we should only consider fermionic descendants. (The symplectic boson descendants in the first line of (\ref{(un)twistedwedgemodes}) all raise the $\cD_0$ eigenvalue by $\frac{1}{2}$.) Let us first analyse the $\cN=4$ case, where the fermionic modes run over the two values 
\be
\psi^i_{\pm \frac{1}{2}} \equiv \psi^i_\pm \ , \qquad 
(\psi^\dagger_I)_{\pm \frac{1}{2}} \equiv (\psi^\dagger_I)_\pm \ . 
\ee
The states at $L_0=0$ (with $\Delta=2$) transform simply as $[0,2,0]$ with respect to the $\mathfrak{su}(4)_R$ symmetry, as follows by the same arguments as in \cite{Gaberdiel:2021jrv,Gaberdiel:2021qbb}. In order to describe them more explicitly, it is convenient to use the DDF operators from eq.~(\ref{DDFold})
\be\label{Sdef}
(\cS_I^{\ i})_0 = \psi^i_+ (\psi^\dagger_I)_- + \psi^i_- (\psi^\dagger_I)_+ \ , \qquad 
(\cS_I^{\ i})_1 = \psi^i_+ (\psi^\dagger_I)_+ \ ,
\ee
that commute with $\cC_0$ and $\cC_1$ since we can discard any modes with mode number bigger than $\frac{1}{2}$ --- they act trivially on the Fock space. Then the physical states at $L_0=0$ can be written as 

\begin{subequations}\label{N4scalarDelta2Worldsheet}
\begin{align}
   &L_0 =0: ~~~ \ket{0}_2 \rightarrow (\mathbf{1}, \mathbf{1})_{-2}~, \qquad \qquad\quad 
        (\cS_1^{\ 3})_0 \, (\cS_1^{\ 4})_0 \, (\cS_2^{\ 2})_0 \, (\cS_2^{\ 4})_0 \,  \ket{0}_2        \rightarrow (\mathbf{1}, \mathbf{1})_{2}~, \label{N4scalarDelta2WorldsheetA} \\
        & ~~~~~~ \ \ \qquad (\cS_I^{\ i})_0 \ket{0}_2 \rightarrow (\mathbf{2}, \mathbf{2})_{-1} \qquad \quad \ \ 
         (\cS_1^{\ 3})_0 \, (\cS_2^{\ 4})_0 \, (\cS_I^{\ i})_0  \ket{0}_2 \rightarrow (\mathbf{2}, \mathbf{2})_{1}~,\label{N4scalarDelta2WorldsheetB} \\
        & ~~~~~~~~\qquad \bigl\{ (\cS_I^{\ i})_0 , (\cS_J^{\ j})_0 \bigr\}\,  \ket{0}_2 \rightarrow (\mathbf{1}, \mathbf{1})_{0}\oplus (\mathbf{3}, \mathbf{3})_{0}~, \label{N4scalarDelta2WorldsheetC}
    \end{align}
\end{subequations}
where we have also spelled out their representation with respect to $\mathfrak{su}(2)_L \oplus \mathfrak{su}(2)_R \oplus \mathfrak{u}(1)_r$. Furthermore, at $L_0=-2$ there is a unique state, namely 
\begin{equation}
L_0=-2:  \qquad (\cS_{1}^{\ 3})_{1}(\cS_{2}^{\ 4})_{1}\, \ket{0}_2 \rightarrow (\mathbf{1}, \mathbf{1})_{0}~.
\end{equation}
This matches exactly the gauge theory spectrum of eq.~\eqref{N4scalarDelta2}.

\subsubsection{$\Delta=2$ states for the $\mathbb Z_{k\geq 3}$ orbifold}

When we apply the $\mathbb Z_{k}$ orbifold, the $\psi^i$ oscillators pick up phases as 
\begin{equation}
    g(\psi^3) = \mathrm{e}^{2\pi\ii/k}\psi^3~,~~~~g(\psi^4) = \mathrm{e}^{-2\pi\ii/k}\psi^4\ . 
\end{equation}
Thus some of the states of the list \eqref{N4scalarDelta2Worldsheet} are projected out since they are not invariant under the orbifold action. In particular, this is the case for the states in \eqref{N4scalarDelta2WorldsheetB}, as well as the six states in \eqref{N4scalarDelta2WorldsheetC} for which $i=j$.  This describes the untwisted sector of the orbifold. The remaining states then come from the different twisted sectors. For example, for the $\mathbb{Z}_3$ case, there are two twisted sectors corresponding to  $\ell= \pm 1$, for which the modes at $w=2$ are, see eq.~(\ref{(un)twistedwedgemodes}) 
\begin{equation}\label{Z3oscillators}
   \begin{array}{lll}
   (\psi^\dagger_I)_{\pm \frac{1}{2}}\ , \quad & (\psi^3)_{\frac{5}{6}}\ ,~~(\psi^3)_{-\frac{1}{6}}\ ,~~(\psi^4)_{\frac{1}{6}}\ ,~~(\psi^4)_{-\frac{5}{6}}  \ , \quad &  \mathrm{for}~ \ell=1~, \\
   (\psi^\dagger_I)_{\pm \tfrac{1}{2}}\ , \quad &   (\psi^3)_{\frac{1}{6}}\ ,~~(\psi^3)_{-\frac{5}{6}}\ ,~~(\psi^4)_{\frac{5}{6}}\ ,~~(\psi^4)_{-\frac{1}{6}} \ , \quad & \mathrm{for}~ \ell=-1~.
\end{array}
\end{equation}
The DDF modes of the form (\ref{DDFopsU}) take the same form as in (\ref{Sdef}), while the `twisted' DDF operators of eq.~(\ref{DDFopsT}) are now explicitly 
\begin{align} \label{Sdeftwis}
({\cal T}_I^{\ 3})_{\frac{\ell}{k}} & = \psi^3_{\frac{1}{2} + \frac{\ell}{k}} (\psi^\dagger_I)_{-\frac{1}{2}} + \psi^3_{-\frac{1}{2} + \frac{\ell}{k}} (\psi^\dagger_I)_{\frac{1}{2}} \qquad ({\cal T}_I^{\ 3})_{1+\frac{\ell}{k}} = \psi^3_{\frac{1}{2} + \frac{\ell}{k}} (\psi^\dagger_I)_{\frac{1}{2}} \\
({\cal T}_I^{\ 4})_{-\frac{\ell}{k}} & = \psi^4_{\frac{1}{2} - \frac{\ell}{k}} (\psi^\dagger_I)_{-\frac{1}{2}} + \psi^4_{-\frac{1}{2} - \frac{\ell}{k}} (\psi^\dagger_I)_{\frac{1}{2}} \qquad ({\cal T}_I^{\ 4})_{1-\frac{\ell}{k}}  = \psi^4_{\frac{1}{2} - \frac{\ell}{k}} (\psi^\dagger_I)_{\frac{1}{2}} \ . 
\end{align}
The states that survive the orbifold projection in the $\ell$-th twisted sector --- the analysis also works directly for $\ell=0$, \textit{i.e.}\ the untwisted sector --- are then 
\begin{equation}\label{N4scalarDelta2Worldsheet1}
    \begin{split}
        &L_0 =0: ~~~ \ket{0}_2 \rightarrow ( \mathbf{1})_{-2}~, \qquad \qquad\quad 
        ({\cal T}_1^{\ 3})_{\frac{\ell}{k}} \, ({\cal T}_1^{\ 4})_{-\frac{\ell}{k}} \, ({\cal T}_2^{\ 3})_{\frac{\ell}{k}} \, ({\cal T}_2^{\ 4})_{-\frac{\ell}{k}} \,  \ket{0}_2        \rightarrow (\mathbf{1})_{2}~, \\
         & ~~~~~~~~\qquad \bigl\{ ({\cal T}_I^{\ 3})_{\frac{\ell}{k}} , ({\cal T}_J^{\ 4})_{-\frac{\ell}{k}} \bigr\}\,  \ket{0}_2 \rightarrow (\mathbf{1})_{0}\oplus (\mathbf{3})_{0}~,  
        \end{split}
\end{equation}
as well as
\begin{equation}
L_0=-2: \qquad ({\cal T}_1^{\ 3})_{1+\frac{\ell}{k}} ({\cal T}_2^{\ 4})_{1-\frac{\ell}{k}}\, \ket{0}_2 \rightarrow (\mathbf{1})_{0} \ . 
\end{equation} 
Thus we get the same set of states for each $\ell$, and these results reproduce exactly what we found in the gauge theory, see eq.~\eqref{Z3transformations}. 

Note that all of these states were created by the `twisted' DDF generators from the spectrally flowed ground state, but in general we shall also need the `untwisted' DDF generators of the form $({\cal U}_M^{\ N})_m$, see eq.~(\ref{DDFopsU}). We have also constructed these states from first principles, \textit{i.e.}\ determined the states of the wedge Fock space that satisfy the residual physical state conditions of eq.~(\ref{physical}). In all cases we have studied, the physical states could always be generated by the action of the untwisted and twisted DDF operators of eq.~(\ref{DDFopsU}) and (\ref{DDFopsT}), and we believe this to be true in general.


\subsubsection{$\Delta=2$ states for the $\mathbb Z_2$ orbifold}

Finally, let us analyse the $\Delta=2$ states for the $\mathbb{Z}_2$ orbifold which behaves slightly different due to the $\mathfrak{su}(2)_L$ symmetry, see the discussion in Section~\ref{sec:4.2}. First of all, it is clear that all the states from (\ref{N4scalarDelta2Worldsheet}), except for \eqref{N4scalarDelta2WorldsheetB}, survive in the untwisted sector, since the $\mathbb{Z}_2$ action just gives a sign for both $\psi^3$ and $\psi^4$. Thus in the untwisted sector we have the $13$ states that we also obtained from the gauge theory perspective, see eq.~\eqref{UntwistedTwistedDelta2}.

In the $\ell=1$ twisted sector, the surviving wedge modes are, see Section~\ref{sec:4.2}
\be
(\psi^\dagger_{I})_{\pm \frac{1}{2}} \ , \qquad \psi^i_{0,1} \ . 
\ee
Now the relevant ${\cal T}$-operators (that commute with $\cC_0$ and $\cC_1$) are 
\begin{align}
({\cal T}_I^{\ i})_{\frac{1}{2}} & = \psi^i_1 (\psi^\dagger_{I})_{- \frac{1}{2}}  + \psi^i_0 (\psi^\dagger_{I})_{\frac{1}{2}}  \\ 
({\cal T}_I^{\ i})_{\frac{3}{2}} & = \psi^i_1 (\psi^\dagger_{I})_{ \frac{1}{2}} \ ,
\end{align}
and thus we get the physical states 
\begin{equation}\label{TwistedDelta2Worldsheet}
    \begin{split}
        &L_0 =0: ~~~ \ket{0}_2 \rightarrow (\mathbf{1},\mathbf{1})_{-2}~,  \\[0.2cm]
        &L_0=-2: ~({\cal T}_1^{\ 3})_{\frac{1}{2}} ({\cal T}_2^{\ 3})_{\frac{1}{2}} \, ({\cal T}_1^{\ 4})_{\frac{1}{2}} ({\cal T}_2^{\ 4})_{\frac{1}{2}}\, \ket{0}_2  \rightarrow (\mathbf{1}, \mathbf{1})_{2}~, \\
        & ~~~~~~~~\qquad  \parenth{({\cal T}_1^{\ (i})_{\frac{3}{2}}({\cal T}_2^{j)})_{\frac{1}{2}}-({\cal T}_1^{(i})_{\frac{1}{2}}({\cal T}_2^{j)})_{\frac{3}{2}}} \ket{0}_2  \rightarrow (\mathbf{3}, \mathbf{1})_{0}  \\ 
        & ~~~~~~~~\qquad \parenth{({\cal T}_{(I}^{\ 3})_{\frac{3}{2}}({\cal T}_{J)}^{\ 4})_{\frac{1}{2}}-({\cal T}_{(I}^{\ 3})_{\frac{1}{2}}({\cal T}_{J)}^{\ 4})_{\frac{3}{2}}} \ket{0}_2 
         \rightarrow   (\mathbf{1}, \mathbf{3})_{0} \\ 
        & ~~~~~~~~\qquad \parenth{({\cal T}_{1}^{\ 3})_{\frac{3}{2}}({\cal T}_{2}^{\ 4})_{\frac{1}{2}}+({\cal T}_{1}^{\ 3})_{\frac{1}{2}}({\cal T}_{2}^{\ 4})_{\frac{3}{2}}} \ket{0}_2 \rightarrow (\mathbf{1}, \mathbf{1})_{0}\ , 
    \end{split}
\end{equation}
which again matches precisely the result from the gauge theory, see eq.~\eqref{UntwistedTwistedDelta2}.

\end{appendix}


\providecommand{\href}[2]{#2}\begingroup\raggedright\endgroup

\end{document}